\documentclass{appolb}
\usepackage{epsfig}
\usepackage{amssymb}
\usepackage{color}

\begin{document}

\title{Criticality Characteristics of Current Oil Price Dynamics}

\author{Stanis{\l}aw Dro\.zd\.z$^{1,2}$, Jaros{\l}aw Kwapie\'n$^{1}$,
Pawe{\l} O\'swi\c ecimka$^{1}$
\address{$^1$Institute of Nuclear Physics, Polish Academy of Sciences,
PL--31-342 Krak\'ow, Poland \\
$^2$Institute of Physics, University of Rzesz\'ow, PL--35-310 Rzesz\'ow,
Poland}}

\maketitle

\begin{abstract}
Methodology that recently lead us to predict to an amazing accuracy the 
date (July 11, 2008) of reverse of the oil price up trend is briefly 
summarized and some further aspects of the related oil price dynamics 
elaborated. This methodology is based on the concept of discrete scale 
invariance whose finance-prediction-oriented variant involves such 
elements as log-periodic self-similarity, the universal preferred scaling 
factor $\lambda \approx 2$, and allows a phenomenon of the "super-bubble". 
From this perspective the present (as of August 22, 2008) violent - but 
still log-periodically decelerating - decrease of the oil prices is 
associated with the decay of such a "super- bubble" that has started 
developing about one year ago on top of the longer-term oil price 
increasing phase (normal bubble) whose ultimate termination is evaluated 
to occur in around mid 2010.
\end{abstract}


\vskip0.5cm

{\it E-mail address: Stanislaw.Drozdz@ifj.edu.pl}

\vskip0.5cm

Recent violent price changes and the related speculative bubbles on the 
world commodity market - such as the precious metals market or the oil 
market - provide further very valuable ground to test the basic components 
of what we globally term "the prediction oriented variant of financial 
log-periodicity"~\cite{drozdz06}. The most relevant of these components 
include the self-similar log-periodicity~\cite{drozdz99} which may 
originate from the discrete scale invariance~\cite{sornette98} and allows 
to make link to the critical phenomena~\cite{sornette03}, the postulated 
existence of the corresponding universal preferred scaling factor $\lambda 
\approx 2$ which is common to all the markets and all time 
scales~\cite{drozdz99,drozdz03}, and even more exotic effects like the 
"super-bubbles"~\cite{drozdz03} - phenomena that seem exclusive to the 
financial markets dynamics.

As far as the commodity market is concerned in a recent 
paper~\cite{drozdz08} (available also as ref.~\cite{drozdzla08}) we have 
presented a prediction for the Spring 2008 reversal of the up trend on the 
precious metals market. Time satisfactorily verified this prediction. 
While ref.~\cite{drozdz08} was in press we started extending the same 
methodology to the oil market. The corresponding log-periodic 
interpretation of the oil price dynamics over the time period 2000-2010 
was publicly disclosed as an insertion (note added) to 
ref.~\cite{drozdzla08} on June 23, 2008 and unchanged is here shown in 
Fig.~1. In fact, already at an earlier stage of these investigations, on 
April 15, 2008, exactly the same illustration has been delivered to 
Wojciech Bia{\l}ek, SEB (Skandinaviska Enskilda Banken) TFI analyst (see 
his website~\cite{bialek}).

\begin{figure}
\begin{center}
\includegraphics[width=0.7\textwidth,height=0.5\textwidth]{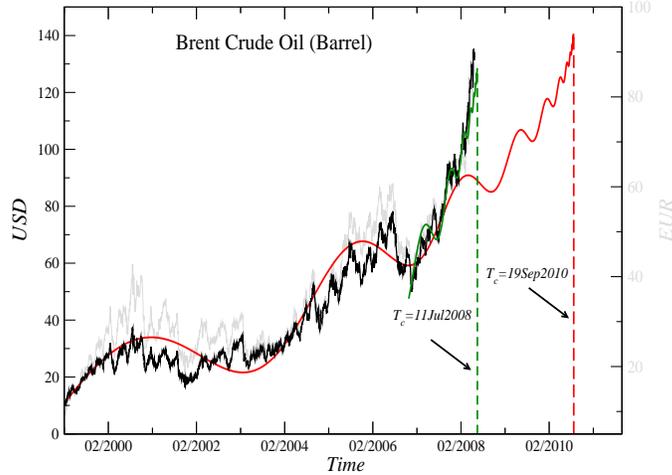}
\caption{This Figure has been posted on the Los Alamos preprint server 
({\it http://axiv.org}) on June 23, 2008 with the following caption: Time 
series of the oil price in USD (grey line represents the oil price in 
Euro) since mid 1999 versus its optimal log-periodic representation with 
$\lambda = 2$. The critical time $T_c$=September 2010 corresponds to an 
ultimate reverse of the long term (since 1999) oil price up trend while 
$t_c$=11.7.2008 sets an upper limit for the end of the present 
"super-bubble".}
\label{fig1}
\end{center}
\end{figure}

Since in the corresponding methodology it is the oscillation pattern that 
carries the most relevant information about the market dynamics, for 
transparency a simple representation of the log-periodicity in the form
\begin{equation}
\Pi(\ln(x)/\ln(\lambda)) = A + B \cos({\omega \over 2\pi} \ln(x) + \phi).
\label{eq:FPE}
\end{equation}
is used, where $\omega = 2\pi / \ln(\lambda)$. The continuous lines seen in
Fig.~1 are then drawn according to the following equation:
\begin{equation}
\Phi(x) = x^{\alpha} \Pi(\ln(x)/\ln(\lambda)), \label{eq:FP}
\end{equation}
where the first term represents a standard power-law that is 
characteristic of continuous scale-invariance with the critical exponent 
$\alpha$. The second term introduces a correction that is periodic in 
$\ln(x)$. In the financial context $x$ represents a distance to the 
critical time $T_c$. Thus $x = \vert T - T_c \vert$, where $T$ denotes the 
clock time labelling the original price time series.

\begin{figure}
\begin{center}
\includegraphics[width=0.7\textwidth,height=0.5\textwidth]{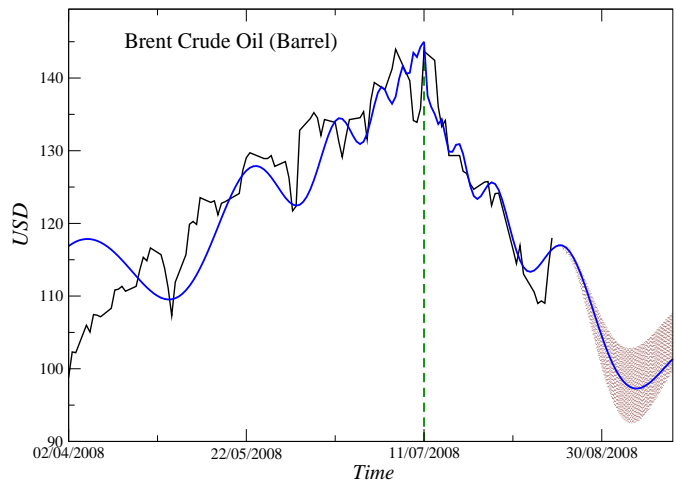}
\caption{Time series of the oil price in USD since 2.4.2008 until 
22.8.2008 (present) versus its optimal log-periodic representation with 
$\lambda = 2$. Accumulation of the log-periodic oscillations on both sides 
of the phase transition date $t_c$ coincide with the previously predicted 
$t_c = 11.07.2008$. The shaded area on the right hand side reflects 
current uncertainty concerning the oil market development in the nearest 
future.}
\label{fig2}
\end{center}
\end{figure}

As is seen in this Figure the sharp increase of the oil price since Summer 
2007 can be recognized to be clearly log-periodic with $\lambda = 2$ 
indicating July 11, 2008 as the end of this phase of increase in the oil 
market. We typically do not demand that much of accuracy but this time the 
agreement was perfect. It was exactly that day when the New York traded 
oil price approached maximum of 148 USD but closed already lower and 
sharply started dropping since. Magnifying the vicinity of this date and 
including the oil price changes up to present (August 22, 2008) results in 
a picture which is illustrated in Fig.~2. A new but consistent element 
seen here is that the present oil market declining phase is developing the 
decelerating log-periodic oscillations starting on July 11 and their 
sequence well corresponds to $\lambda = 2$ - a condition which makes it 
reliable. Accordingly, extrapolation of this oscillatory pattern points to 
the possibility of a further oil price decline in the coming weeks to a 
level even below 100 USD. Such a possibility has already been expressed in 
ref.~\cite{drozdzla08}. After that however, from what we learn~\cite 
{bartolozzi} from the stock markets in analogous phases of their dynamics, 
such a log-periodically decelerating phase is likely to dissolve and thus 
the market may start resuming the up trend.

\begin{figure}
\begin{center}
\includegraphics[width=0.7\textwidth,height=0.5\textwidth]{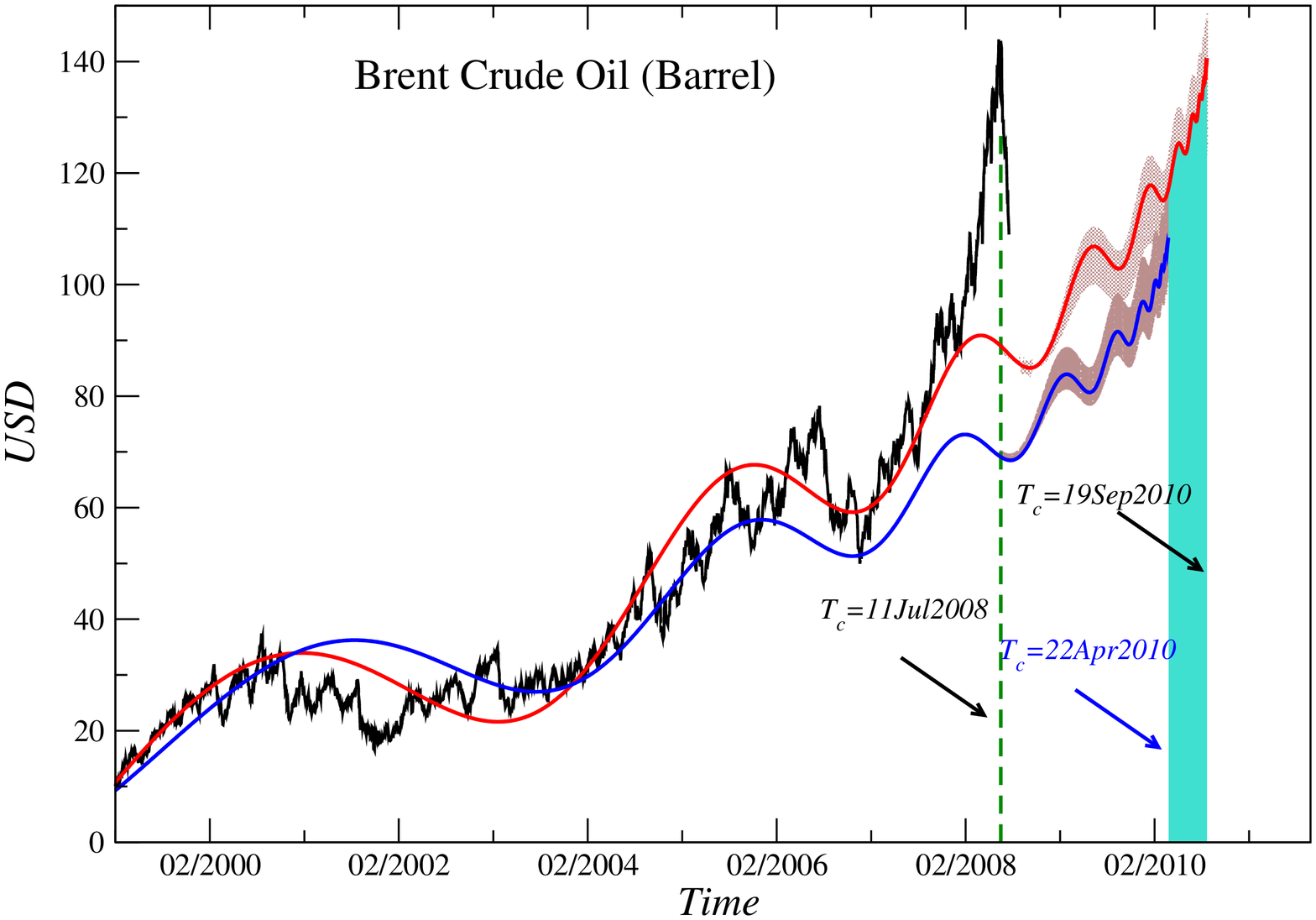}
\caption{Time series of the oil price in USD since mid 1999 until August 
22, 2008 versus two extreme log-periodic representations with $\lambda = 
2$ that still may correlate with the oil market dynamics. Shaded areas 
reflect current uncertainties regarding both $T_c$ and the amplitude of 
price changes.}
\label{fig3}
\end{center}
\end{figure}

Such a scenario is also in accord with the longer time perspective to view 
the oil price dynamics. As it is also shown in Fig.~1, in order to grasp 
the corresponding development since about the year 2000 within the 
consistent log-periodic framework with the same preferred universal 
scaling factor $\lambda = 2$ the last months oil price accelerated 
elevation is to be considered as one of the consecutive increases in the 
sequence of long-term log-periodic pattern which gets boosted into a local 
bubble on top of a long-term bubble and therefore it is termed a 
"super-bubble"~\cite{drozdz03}. History of the financial markets provides 
several examples of such effects~\cite {drozdz03}. When such a 
"super-bubble" crashes the system returns to a normal bubble state that 
eventually ends at the time determined by the long-term patterns. In the 
present case of the oil market, as it has been indicated in Fig.~1, the 
time of un ultimate reverse of the present long-term up trend may 
correspond to the late Summer in 2010, thus several months after the stock 
market enters a serious recession. A more precise evaluation of the 
related $T_c$ should be possible after the oil market fully expresses the 
size and duration of the correction due to decay of the recent 
"super-bubble" that has ended on July 11, 2008. The current degree of 
uncertainty regarding $T_c$ is indicated in Fig.~3.

Interpreting thus recent developments on the world oil market in terms of 
the financial log-periodicity provides further arguments in favor of this 
theoretical concept and, in particular, of such its elements like the 
existence of the universal preferred scaling factor $\lambda \approx 2$ 
and appearance of the "super-bubbles". The precision of the reported here 
real prediction exposes a predictive potential of the corresponding 
methodology.

\vspace{1cm}

We thank Wojciech Bia{\l}ek for a stimulating correspondence.

\end{document}